# Geometric interference in a high-mobility graphene annulus *p-n* junction device

Son T. Le[1,2,3], Albert F. Rigosi[1], Joseph A. Hagmann[1], Christopher Gutiérrez[1,4,5], Ji Ung Lee[2], Curt A. Richter[1*]

[1]*Physical Measurement Laboratory, National Institute of Standards and Technology (NIST), Gaithersburg, MD 20899, United States*

[2]*College of Nanoscale Science and Engineering, State University of New York Polytechnic Institute, Albany, New York 12203, United States*

[3]*Theiss Research, Inc., La Jolla, CA 92037, United States*

[4]*Maryland NanoCenter, University of Maryland, College Park, Maryland 20742, United States*

[5]*Department of Physics and Astronomy, University of California, Los Angeles, Los Angeles, CA 90095, United States*

ABSTRACT: The emergence of interference is observed in the resistance of a graphene annulus *pn* junction device as a result of applying two separate gate voltages. The observed resistance patterns are carefully inspected, and it is determined that the position of the peaks resulting from those patterns are independent of temperature and magnetic field. Furthermore, these patterns are not attributable to Aharonov-Bohm oscillations, Fabry Perot interference at the junction, or moiré potentials. The device data are compared with those of another device fabricated with a traditional Hall bar geometry, as well as with quantum transport simulation data. Since the two devices are of different topological classes, the subtle differences observed in the corresponding measured data indicate that the most likely source of the observed geometric interference patterns is quantum scarring.

---

[*] Email: curt.richter@nist.gov

# I. INTRODUCTION

Graphene exhibits unique properties [1-4], and graphene-based devices featuring hexagonal boron nitride (*h*-BN) as an encapsulation and support layer show an enhanced level of these properties, as well as other interesting phenomena [5-8]. A few examples of such phenomena include the observation of moiré superlattices [9-11], Hofstadter's butterfly [12-16], the quantum Hall effect in *p-n* junctions (*pn*Js) [17-30], and, most relevantly, quantum scarring [31-35]. These high mobility graphene-based devices and other similar devices that host systems dependent on Fermi-Dirac statistics have become a valuable materials platform for exploring two-dimensional (2D) physics and can additionally be applied towards photodetection [36-40], quantum Hall resistance standards [41-45], and electron optics [46-48].

Annulus *pn*J devices appear in various facets of research, including investigations of Aharonov-Bohm (A-B) oscillations and mesoscopic valley filters [49-52]. Coupled with the exfoliation of *h*-BN, these graphene devices have their qualities enhanced [5, 6, 13, 51], leading to observations of quantum scarring [31, 32]. However, there has been little exploration in the realm of electronic interference in graphene annulus *pn*J devices, which can be made to host a bipolar charge carrier population.

In this work, the emergence of patterns was observed in the resistance of a graphene annulus device as a function of gate voltage. The substrate contains two local, embedded gates below the device, enabling the formation of a *pn*J that separates both halves of the annulus. Analysis of the observed resistance patterns ($\Delta R$) reveal peak positions within the patterns that are independent of temperature and invariant to broken time-reversal symmetry. These patterns are not attributable to A-B oscillations, but rather arise due to the geometry of the device. To show this, we compare data from the device with annulus geometry with those of another device fabricated with a traditional Hall bar geometry. Since the two devices are of different topologies, differences in the corresponding data suggest that the most likely source of the observed geometric interference patterns is quantum scarring. Measurement data were also compared with results from quantum transport simulations performed with the *Kwant* software package.

## II. EXPERIMENTAL AND NUMERICAL METHODS

### A. Sample Preparation

A heterostructure device based on graphene and *h*-BN was assembled by using the flake pick-up method [51]. Standard electron beam lithography and reactive ion etching processes were used for device fabrication. In Fig. 1 (a), an atomic force microscope image shows the top view of the device, with two cross-section profiles showing the *h*-BN/graphene/*h*-BN stack on top of the Si/SiO$_2$ substrate (Fig. 1 (b) and (c)). The substrate has two local, embedded gates below the surface formed from poly-Si, represented by blue and red in Fig. 1 (a) [21]. The gates were atomically smoothed by chemical and mechanical polishing, with a separation between the buried gates of about 100 nm. The designed inner radius is 250 nm, whereas the designed outer radius is 500 nm. The two buried gates have a depth of about 150 nm. The contact terminals and back gates are numerically labelled in Fig. 1 (a). A standard Hall bar was also assembled for comparison and to verify simulation results, with similar fabrication conditions being applied. The atomic force microscope (AFM) image in Fig. 1 (a) was acquired in tapping mode with a scan rate of 1 Hz and has a scan size of approximately 3.5 µm x 3.5 µm.

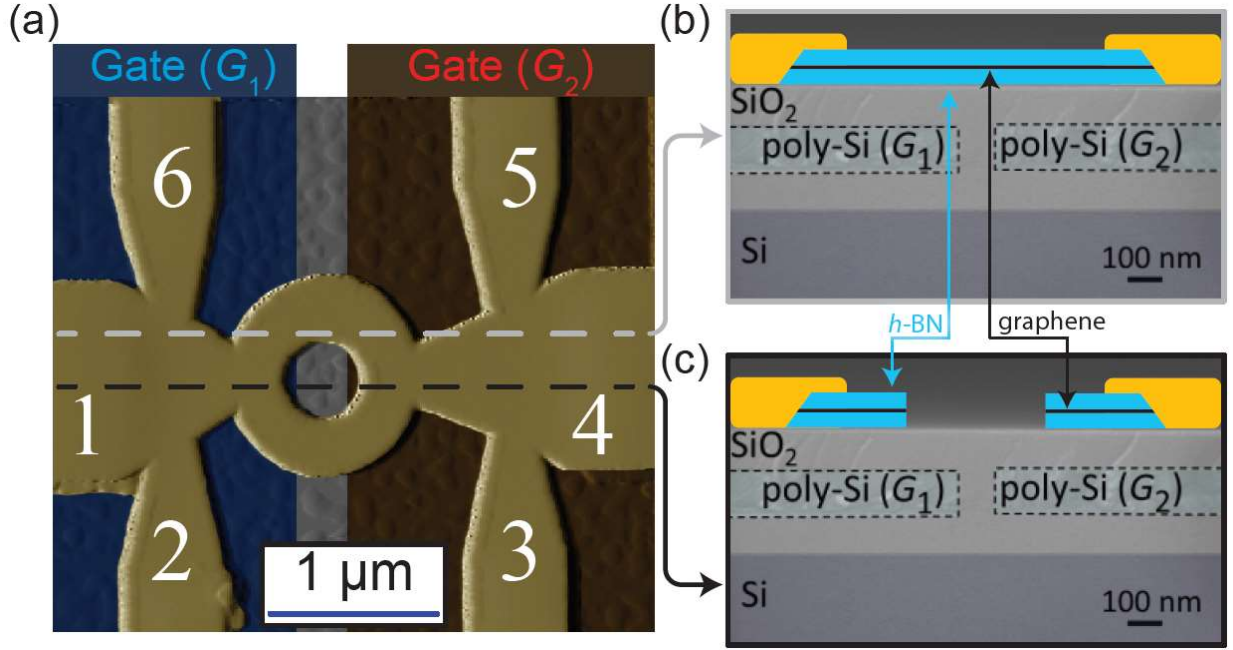

FIG. 1. (Color online) Edge state test structure. (a) An AFM image of the device is shown, with labels indicating the number of the electrical contact. Two buried gates, $G_1$ and $G_2$, are represented by blue and red shading, respectively. The green and purple dashed lines mark cross-sectional images overlaid with illustrative elements in (b) and (c).

## B. Low-Temperature Electrical Transport

The data were collected at zero-field, but strong magnetic fields were used to characterize the quantum Hall properties of the device to verify typical functionality. These data are presented in the Supplemental Material [53]. Transport measurements were performed between 0.3 K and 30 K, as well as between 0 T and 12 T. Traditional lock-in amplifier techniques were used along with currents ranging from 5 nA to 50 nA at 19 Hz. The estimated mobility was 40,000 V cm$^{-2}$ s$^{-1}$ for a carrier density of $10^{12}$ cm$^{-2}$ and 200,000 V cm$^{-2}$ s$^{-1}$ for a carrier density of $10^{10}$ cm$^{-2}$. Both graphene devices had buried gates with which to tune the *pn*Js. An extensive analysis of the expected magnetic-field dependent behavior was explored in previous reports, all using tunable gates to adjust the *pn*J [17, 19, 21, 23]. The longitudinal resistance (and Hall resistivity in the Supplemental Material [53]) was measured as a function of both applied gate voltages ($G_1$ and $G_2$), yielding a 2D parameter space, or map, of the resistance. Various models of quantum transport in these devices were implemented using the *Kwant* package [54].

### III. DEVICE CHARACTERIZATION

The first general characterization measurement involved current injection through contacts 1 and 4, with the voltage measured across contacts 2 and 3 (the bottom *pn*J). The resistance profile was determined as a function of the two gate voltages (at zero-field and 1.8 K), where each gate is buried beneath separates halves of the device (as seen in Fig. 1 (a)). An example resistance profile along $V_{G1} = V_{G2}$ was extracted and shown in Fig. 2 (a). The Dirac point of the device occurs at approximately -2.6 V and serves as a rough center for all voltage plots. Some examples of resistance patterns are indicated by black arrows.

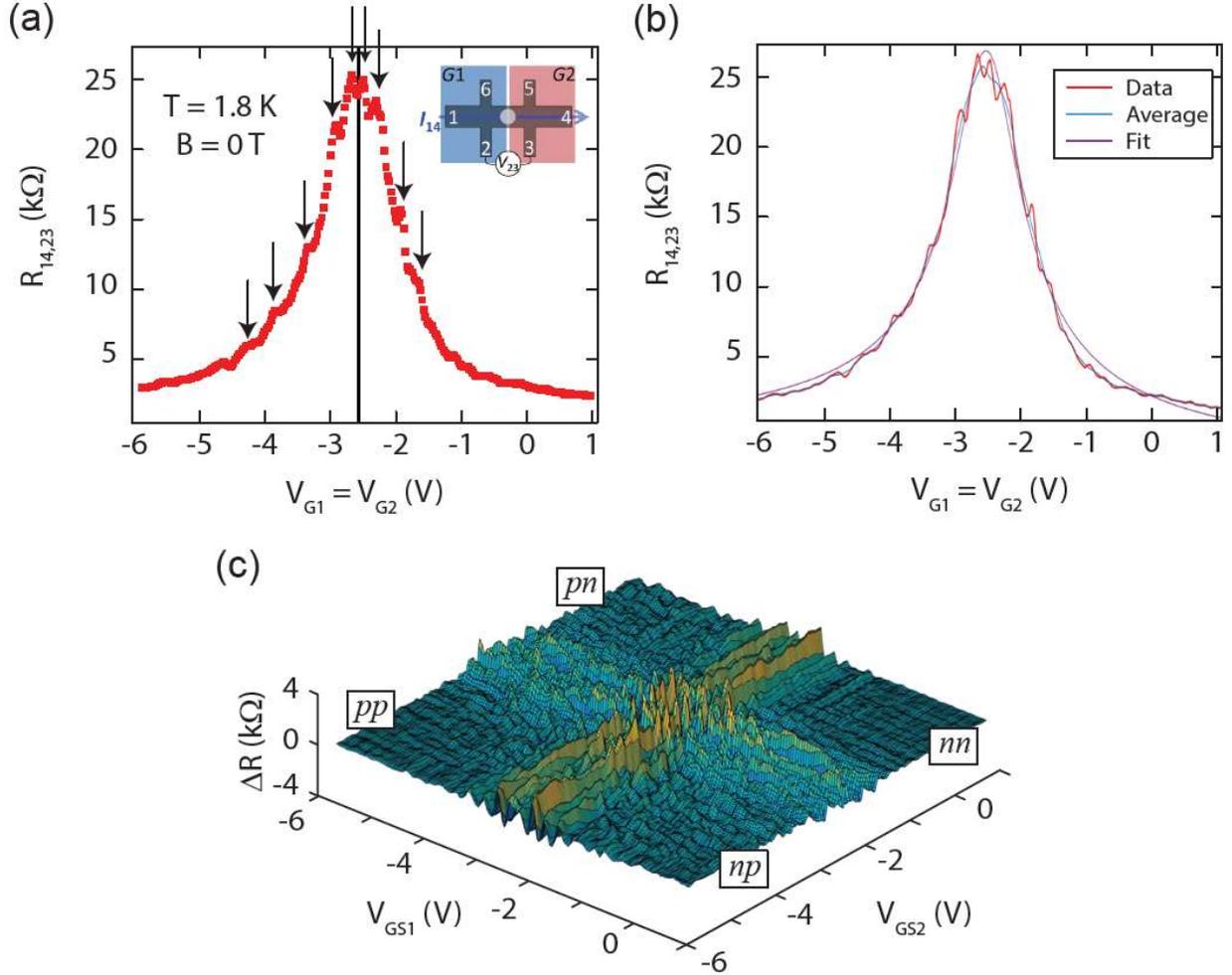

FIG. 2. (Color online) Resistance of test structure. (a) The resistance profile along $V_{G1} = V_{G2}$ (as defined in Fig. 1 (a)) is shown (with the Dirac point occurring at approximately -2.6 V). The patterns in resistance are indicated by black arrows. (b) Two types of fits are performed for data analysis. The purple curve represents the Dirac peak fitting for mobilities and carrier density extraction (labelled "Fit"), whereas the blue curve is a smoothed background used to enhance the nature of the patterns. (c) The smoothed curves are subtracted from both gate voltages to enhance the $\Delta R$ map. The types of unipolar ($n/n$, $p/p$) or bipolar ($p/n$, $n/p$) regions are shown in each corner of the map.

Since the patterns are much smaller than the main resistance peak typical of zero-field measurements, a smoothed fitted background, obtained by an adjacent averaging of the data points for a window of 10 points, was subtracted from all the data (blue curve in Fig. 2 (b)) to enhance the nature of the patterns. A second type of fit was applied by means of a fitting function (purple Lorentzian curve in Fig. 2 (b), labelled "Fit") to extract values of mobility and carrier density. After the full background was subtracted from all the data, a 3D plot, shown in Fig. 2 (c), more clearly displayed the $\Delta R$ as a function of both gates.

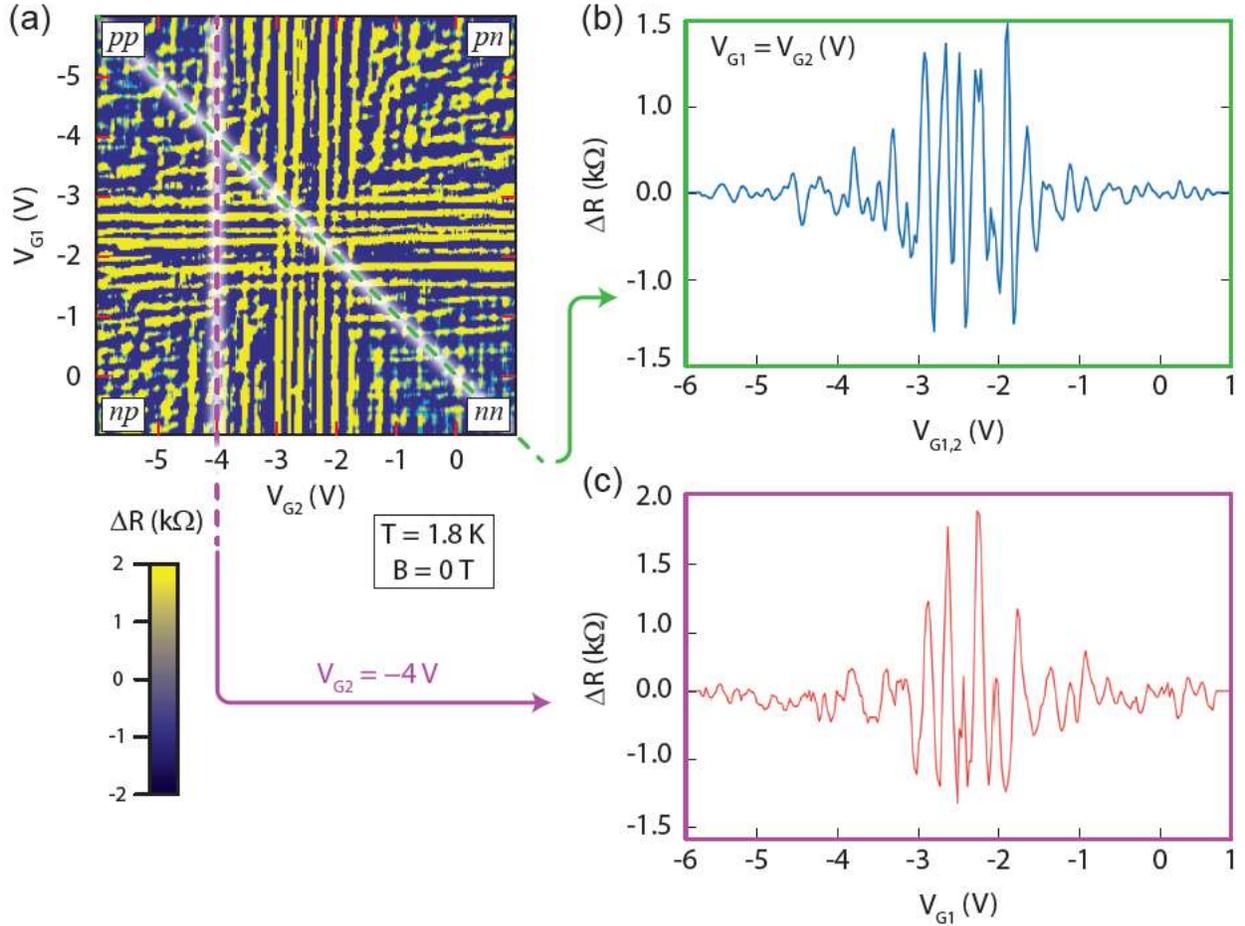

FIG. 3. (Color online) (a) A 2D projection of the $\Delta R$ map of Fig. 2 (d) is shown to clarify which profiles are extracted for the discussion of the data analysis. (b) The first profile is taken along the unipolar diagonal of the map, with the profile exhibiting a strong oscillatory behavior near the Dirac point. (c) Similar behaviors are seen for the second profile taken along $V_{G2} = -4\,V$.

To simplify the analysis, a 2D projection of the $\Delta R$ map in Fig. 2 (d) was generated to clarify the example profiles that were extracted for the discussion of the data analyses, as seen in Fig. 3 (a). The first profile, represented by a dashed green line in Fig. 3 (b), was taken along the unipolar diagonal of the map. The profile shows a strong oscillatory behavior when both halves of the device are near the Dirac point. The amplitude of the patterns reached as high as 1.5 k$\Omega$ and maintained at least 10 % of its maximum value for a few volts. A second profile, represented by a purple dashed line, was taken at $V_{G2} = -4\ V$ and is shown in Fig. 3 (c). Since this profile is not on a diagonal, the pattern amplitude diminished more rapidly with voltage than the diagonal profile.

## IV. RESISTANCE PATTERN ORIGINS

### A. Aharonov-Bohm and Geometric Considerations

The data in Fig. 3 (a) show an oscillatory behavior in the measured resistance of the device as one or both of the two buried gates induces a near-Dirac-point doping in the graphene/$h$-BN heterostructure. To supplement the understanding of these patterns, additional possible dependencies were investigated. For instance, in Fig. 4 (a) and (b), the same resistance map was measured at 4.1 K and 310 mK, respectively, with no strong evidence of temperature dependence aside from the sharpening of local features. The sharpening of local features is evident when comparing the unipolar regions to the bipolar regions, where the latter regions appear to exhibit stronger oscillations. A possible reason for the asymmetry appearing suppressed at lower temperatures is that the electron phase coherence and mobility are improved with lower temperature. This improvement may sharpen features otherwise concealed by higher temperatures. In the bipolar regions, the coherence and mobility conditions need only apply in half of the annulus since the junction would modify the electron path and behavior. This behavior is later support by simulations (see bipolar diagonal behaviors in Fig. 5), namely by the presence of a sturdier electrical response in the bipolar regions. The main central features remained visible up to 30 K.

A magnetic field was used to break time-reversal symmetry, to determine these maps' field-dependence, and to uncover possible contributions of A-B oscillations. Shown in Fig. 4 (c) and (d), the measurements were performed at 5 K using zero-field and 0.1 T, respectively, with evidence that the interference pattern persists even with low magnetic fields. An example measurement of differential resistance is shown in Fig. 4 (e), where A-B oscillations were extracted from the data between 0 T and 0.1 T. When magnifying the region shown in light blue, the periodicity of the A-B oscillations became straightforward to determine. Furthermore, the amplitude of the Fourier transform of these patterns is plotted as a function of inverse magnetic field in Fig. 4 (f). We note here that the spread of the peaks is consistent with the range of areas of the annulus (that is, the area between the two radii). The plot shows a distribution that can be fit with a single peak representing $h/e$. More specifically, this peak is located at about 89 $T^{-1}$ and corresponds to an A-B period of approximately 11 mT. Using the formula $\pi R^2_{avg} \Delta B = h/e$, $R_{avg}$ is calculated to be about 340 nm, which is consistent with the actual average radius of the device (outer radius of 500 nm and inner radius of 175 nm). The slightly larger spread of the $h/e$ peak also indicates that this device is not perfectly ballistic.

Due to an underwhelming amplitude compared with those observed in the resistance maps (of order 10 Ω and 1 kΩ, respectively), A-B oscillations are not thought to be the dominant contribution. Thus, it is possible that device geometry and configuration have significant effects on the observed patterns in $\Delta R$. This second consideration was investigated by examining the results of various simulations using *Kwant*. These numerical simulations were performed to predict observed resistance patterns in transport measurements across the *pn*Js. A tight-binding model was used for a 2D system composed of a graphene layer in the shape of an annulus, as well as a second model for the case of a filled circle.

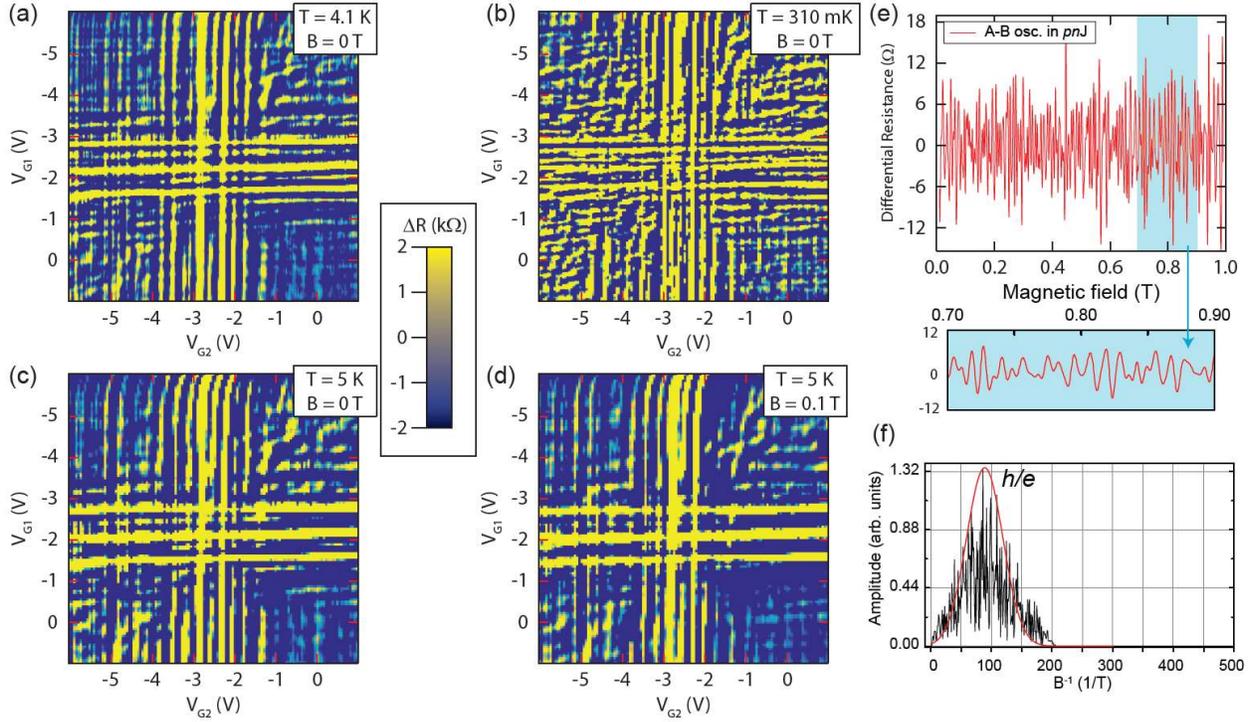

FIG. 4. (Color online) The Δ$R$ maps are shown with varying parameters to show temperature and magnetic field dependence. The analysis done here is meant to determine if Aharonov-Bohm (A-B) oscillations have any significant contribution to observations in the resistance maps. The maps are taken at zero-field and (a) 4.1 K, (b) 310 mK, and (c) 5 K, with (d) being taken at 5 K and 0.1 T. Note that the case of 1.8 K is shown in Fig. 3. All gate voltage graphs have axes that are vertically reflected to match simulation axes. (e) A-B oscillations were extracted from the data between 0 T and 0.1 T, with a magnified region shown just below the graph in light blue to determine the periodicity of those oscillations. (f) The Fourier transform of the amplitude of these oscillations is plotted to determine the distribution seen in the data. This distribution was fitted with one peak representing *h/e*.

In short, the programs methodologically defined the graphene lattice, circular scattering regions, and width and potentials related to the *pn*J. The hopping parameters in the graphene lattice were defined for both types of scattering regions (hopping term is -1 whereas the site potential is ±1 depending on the next neighboring atom), as were the leads, enabling a full calculation of some eigenvalues of the closed system. Specific assumptions include no crystal defects in the graphene and a disordered junction potential as one goes from *p*-type to *n*-type regions (the randomness parameter for the disordered junction is 0.1).

Ultimately, we consider three possible effects: (1) Fabry-Perot interference across the buried gates, (2) patterns from moiré pattern effects on transport, and (3) quantum scarring from a mesoscopic, circular device geometry.

The first set of simulations was carried out to model contributions from Fabry-Perot interference. In the actual device (topologically depicted in Fig. 5 (a) and experimentally realized as a conventional Hall bar device), each half has a buried gate beneath it, with the lateral spacing between the two buried gates shown in Fig 1. These conditions can be simulated by defining the carrier density in each half, which is identical to specifying a gate voltage. With two different gate voltages now defined, the expected voltage $V_{23}$ is simulated (whose experimental counterpart can be seen in Fig. 5 (d) in the upper right inset). These simulated curves are fit in a similar way to Fig. 2 (b) to get a predicted $\Delta R$ for each pair of defined gate voltages. When compiled for all gate pairs, the simulations form the map shown in Fig 5 (b).

From the $\Delta R$ map in Fig. 5 (b), it is not clear that a circular geometry results in the interference observed in Fig. 3 (a). In fact, the map appears dim in comparison to Fig. 3 (a), suggesting that, in this case, a circular geometry diminishes the observed patterns. To grant some validity to the simulation, a topologically equivalent device — one with no missing region in the center: a conventional Hall bar — was fabricated and measured, yielding the data in Fig. 5 (c). These data offer validity to the model since no substantial patterns were observed in the $\Delta R$ map in Fig. 5 (c), with a sample cut (black curve) shown in Fig. 5 (d) compared with a transparent blue overlay of the pattern from Fig. 3 (b). Due to the significant difference of the two curves, one may say Fabry-Perot interference is not a significant contributor to the patterns and suggests that the topological class of the device was of importance for understanding these observations.

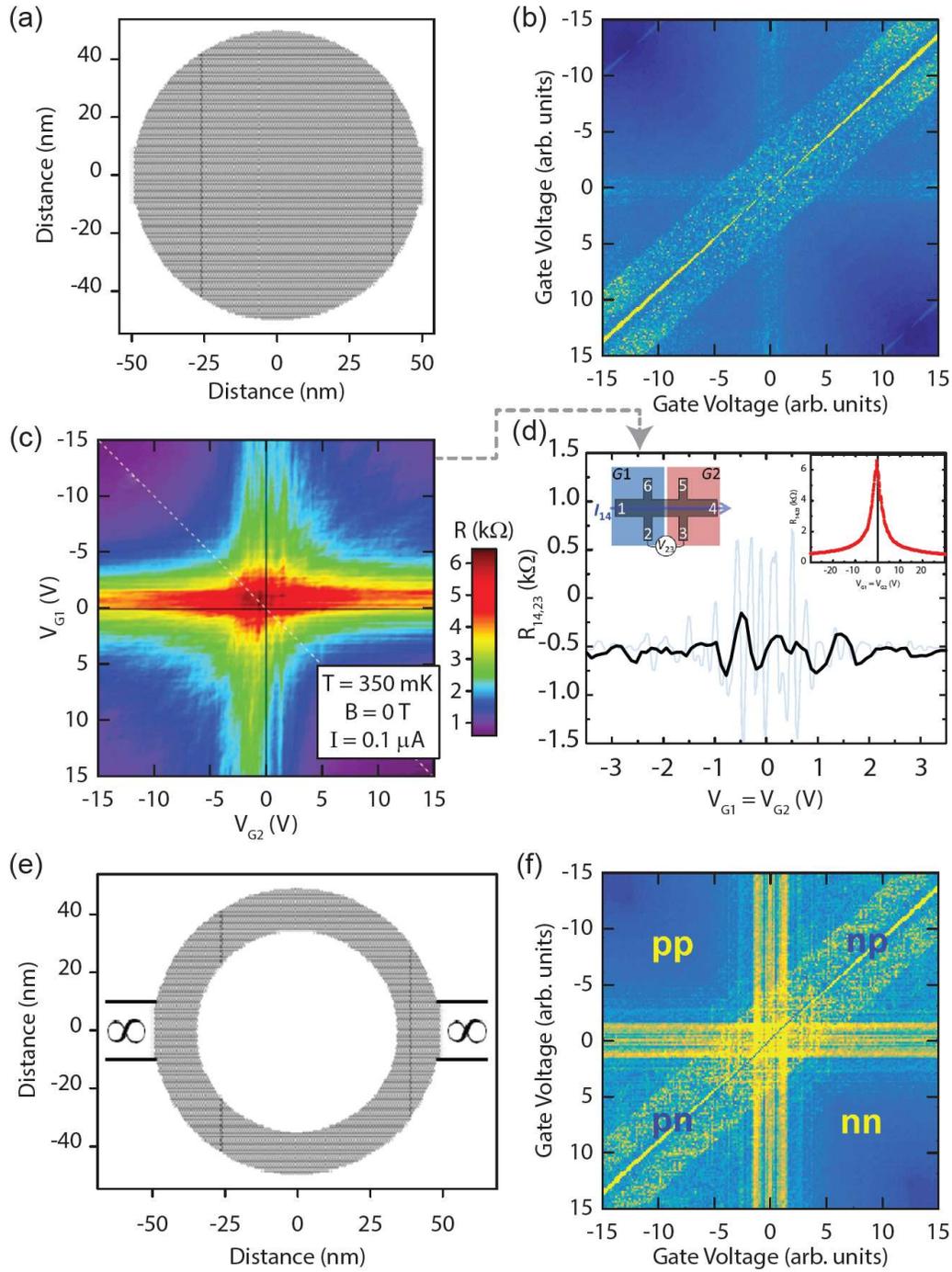

FIG. 5. (Color online) (a) The software package *Kwant* was used to simulate Δ*R* maps in devices similar to the one fabricated in Fig. 1. The first simulation was for a graphene circle of similar outer radius to the device. (b) The simulated Δ*R* map for (a) is shown here, bearing little resemblance to the observed behavior. (c) A total resistance profile is measured on a standard

Hall bar device, which shares a topological resemblance to the simulated system in (a). No significant patterns are observed in this case, indicating that the missing central region is crucial to the simulation rather than the circular shape. An example curve along the unipolar diagonal in dotted white. (d) The example curve (red in upper right inset) has its background subtracted in the same manner as Fig. 2 (c) and is plotted (black curve) alongside the curve from Fig. 3 (b) (transparent blue curve) to highlight the lack of periodicity. (e) The corresponding model is shown for a graphene device in the shape of an annulus. (f) The simulated $\Delta R$ map for the annulus bears a more similar resemblance to the data, namely in its display of electronic interference near the Dirac point. All gate voltage graphs have axes that are vertically reflected to match simulation axes.

### B. Quantum Scarring

After validating the simpler model, we adjusted the model for the second topological class: the annulus-shaped device (as described in Fig. 1). A finite region with an identical aspect ratio was used and was coupled to infinite leads, as seen in Fig. 5 (e). The resulting predicted $\Delta R$ map in Fig. 5 (f), which only accounted for device geometry, appeared to be very similar to our experimental results. Again, because the maps from Fig. 5 (b) and 5 (f) are not similar, despite the similar gating conditions, Fabry-Perot interference is not thought to be a large contributing factor to the experimental observations.

In essence, *Kwant* allows us to predict a matching result without necessarily revealing the mechanism behind the observation, and this is why we require this process of elimination. Before discussing the other possible contributing factors for these observations, one should note that the bipolar diagonal feature in both Fig. 5 (b) and (f) is partly associated with the width of the wires making up the simulated annulus.

The second possible contribution could be from modifications in the graphene band structure from the moiré pattern formed in the device from stacked *h*-BN. This contribution was not modelled with *Kwant*, but it has been reported that electron transport in graphene can be affected by a moiré pattern potential [27], especially if their crystallization orientations are aligned within 1°, giving a moiré wavelength larger than 10 nm. Since the experimental devices were intentionally misaligned by tens of degrees, and by comparing these observations with those predicted in Ref. [27], it becomes clear that the characteristic patterns in resistance seen here do not reflect the expected effects from moiré potentials, most notably in the lack of the prominent satellite Dirac peaks. Since moiré potentials are not expected to contribute heavily to $\Delta R$, this eliminates this second possible effect as a contributor to the observed $\Delta R$ pattern.

The third possible contributor is quantum scarring, which has been discussed in the literature [31-35]. In short, the device geometry and lateral dimensions are within a small window of allowable conditions such that wave functions describing the crossing charge carriers may develop small pockets or regions where the particle probability densities are predicted to be higher or lower [55-60]. Since graphene is a relativistic Dirac system [34], the periodicity of the scar patterns varies linearly with the Fermi energy contra the case of a conventional semiconductor system, which exhibits a periodicity that varies with the square root of the Fermi energy [31]. Another interesting observation to note comes from a recent work in which similar devices that are topologically similar to the measured conventional Hall bar also do not exhibit quantum scarring [61]. Recall that, in Fig. 5 (a), simulations for simply connected regions are found to not have any significant contribution to the observed resistance patterns.

In order to investigate the possible extent of quantum scarring, the same models were used to calculate the local density of states (LDOS) within the annulus. Three cases were calculated and shown in Fig. 6 (a), (b), and (c). In the first case, the LDOS of a device with unipolar doping exhibits some symmetric scarring. In the second case (Fig. 6 (b)), one half of the device is maintained near the Dirac point, and, consequently, the interference reduces overall on the charged half while nearly vanishing in the neutral half.

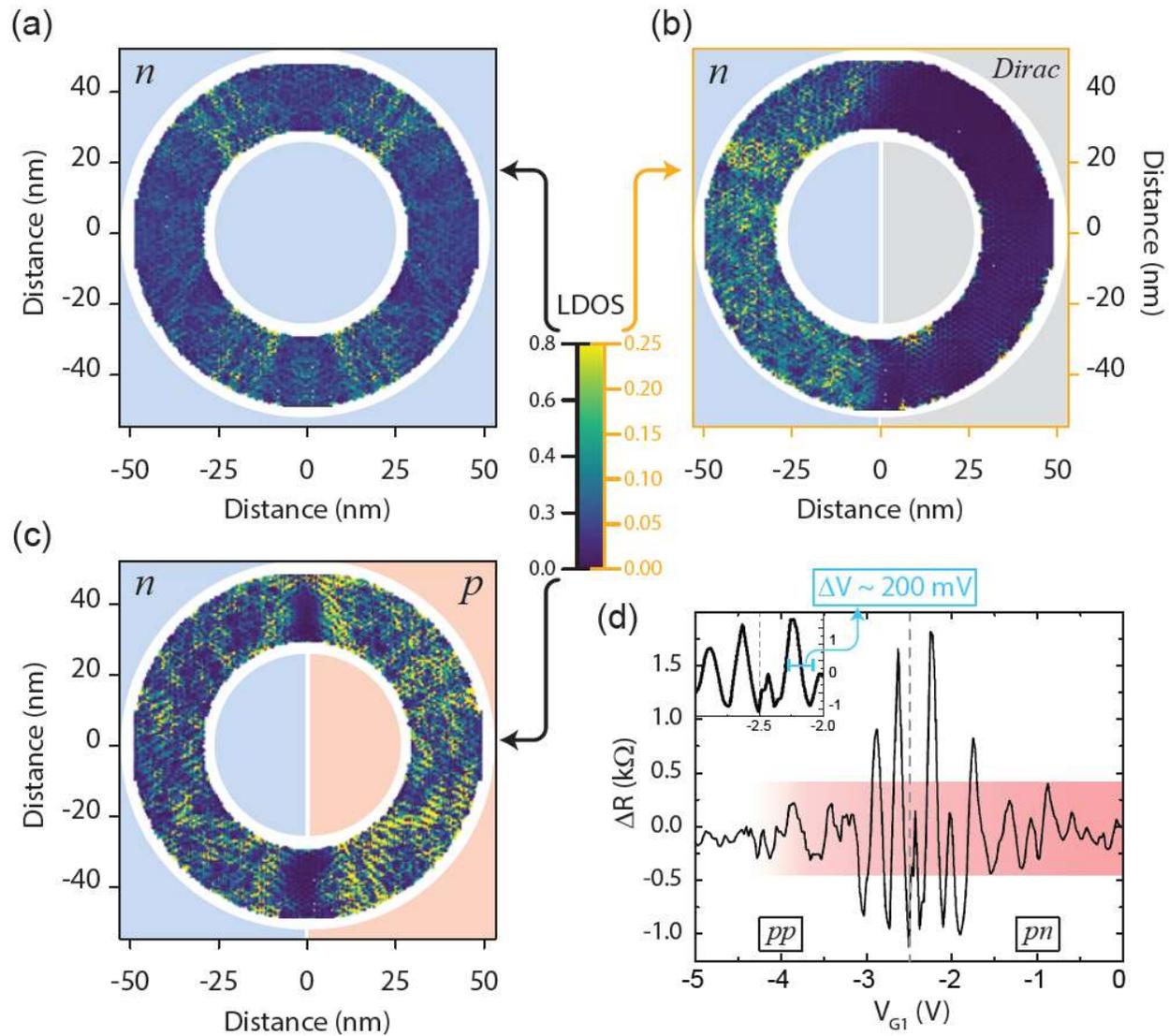

FIG. 6. (Color online) (a) The local density of states (LDOS) is calculated for the geometry of the actual device (unipolar case). (b) The LDOS is calculated for the case where the right half of the device is at the Dirac point and (c) when that same region is tuned such that the device becomes bipolar. The left axis of the color scale applies to (a) and (c) whereas the right axis

(orange) applies to (b) only. (d) A $\Delta R$ profile is used here to demonstrate the inherent asymmetry in amplitude between the unipolar and bipolar configurations. Further, the profile's periodicity is extracted to compare with other observations in the literature. All polarities in these calculations were set to $\pm 10^{12}$ cm$^{-2}$.

In the third case, shown in Fig. 6 (c), a bipolar doping arrangement was simulated, and an asymmetry in the LDOS over the geometry of the device emerged. There are two points of support that can be made to justify at least a partial attribution of our observations to quantum scarring. The first point comes from the inherent asymmetry about the Dirac point in the extracted profiles of the $\Delta R$ map. The profile shown in Fig. 6 (d) is the measurement taken at 1.8 K (zero-field) while $V_{G2}$ is held at -4 V. The data show a consistent asymmetry in how rapidly the amplitude of $\Delta R$ decreases. As the gate voltage brings the device from unipolar to bipolar, the amplitude of the pattern decreases in amplitude more slowly after crossing the Dirac point. This is indicated in Fig. 6 (d) by using a red-colored region to mark the range of the larger pattern amplitudes (present in a bipolar device). For a case like Fig. 3 (b), the diagonal profile does not clearly exhibit this asymmetry since the Dirac point is separating two unipolar doping regimes. This asymmetry is also visible in the overall 2D $\Delta R$ map like the one in Fig. 4 (a). It is clear from the data that the patterns have a greater intensity in the bipolar regimes (lower left and upper right quadrants of the graph) than in the unipolar ones.

The asymmetry in the bipolar regime of the $\Delta R$ map then becomes correlated with the increased LDOS that arises, as seen in Fig. 6 (c). The greater LDOS seen in the bipolar arrangement may suggest a greater magnitude of conductivity, especially when compared to the LDOS in the unipolar case (Fig. 6 (a)). The second point of support that the observed effect may be attributable to quantum scarring involves the periodicity of patterns. By magnifying the region close to the Dirac point in Fig. 6 (d), one can see that the period of $\Delta R$ is approximately 200 mV.

Further analysis of this example curve, through taking the Fourier transform of the data, reveals that a more precise calculation of the periodicity is 219 mV ± 52 mV. To help gauge whether this is a reasonable periodicity, this result is compared with another observed periodicity in a similar unipolar device. Since the devices in Ref. [31] exhibit a similar periodicity in the mid-100s of mV, this further suggests that quantum scarring may be the most likely dominant contributor to the effect observed in $\Delta R$.

## V. CONCLUSIONS

In this work, the emergence of geometric interference was observed by means of measuring the longitudinal resistance of a graphene annulus device as a function of two gate voltages. The resistance patterns were determined to be independent of temperature and magnetic field. These patterns were not attributed to Aharonov-Bohm oscillations, but rather are shown to most likely arise due to the geometry of the device, which enables the phenomenon of quantum scarring to occur. Quantum scarring predictions were made by simulating various device geometries with the *Kwant* software package. Observations of resistance pattern asymmetry in the data served as one point of support for quantum scarring, with the second point of support coming from the assessment of the periodicity of the resistance patterns and a comparison to similar observations in the literature. In conclusion, it is likely that inference due to quantum scarring in this device geometry is a dominant contributor to the observed resistance patterns near the Dirac point for graphene annulus *pn* junction devices.

## ACKNOWLEDGMENTS AND NOTES

S.T.L. acknowledges support from the U.S. Department of Commerce, National Institute of Standards and Technology under the financial assistance award 70NANB18H155. C.G.

acknowledges support under the Cooperative Research Agreement between the University of Maryland and the National Institute of Standards and Technology, Grant No. 70NANB10H193, through the University of Maryland. The authors thank all those involved with the internal NIST review process. Commercial equipment, instruments, and materials are identified in this paper in order to specify the experimental procedure adequately. Such identification is not intended to imply recommendation or endorsement by the National Institute of Standards and Technology or the United States government, nor is it intended to imply that the materials or equipment identified are necessarily the best available for the purpose. The authors declare no competing interest.

Supplemental Material

# Geometric interference in a high-mobility graphene annulus *p-n* junction device

Son T. Le[1,2,3], Albert F. Rigosi[1], Joseph A. Hagmann[1], Christopher Gutiérrez[1], Ji U. Lee[2], Curt A. Richter[1†],

[1]*Physical Measurement Laboratory, National Institute of Standards and Technology (NIST), Gaithersburg, MD 20899, United States*

[2]*College of Nanoscale Science and Engineering, State University of New York Polytechnic Institute, Albany, New York 12203, United States*

[3]*Theiss Research, Inc., La Jolla, CA 92037, United States*

**Quantum Hall Transport**

The data in the main text were collected at zero-field, but strong magnetic fields were used to characterize the quantum Hall properties of the device to verify typical functionality. These data are presented in Fig. 1-SM. Transport measurements were performed between 0.3 K and 30 K, as well as between 0 T and 12 T. The estimated mobility was 40,000 $Vcm^{-2}s^{-1}$ for a carrier density of $10^{12}$ $cm^{-2}$ and 200,000 $Vcm^{-2}s^{-1}$ for a carrier density of $10^{10}$ $cm^{-2}$. Both graphene devices had buried gates with which to tune the *pn*Js.

---

[†] Email: curt.richter@nist.gov

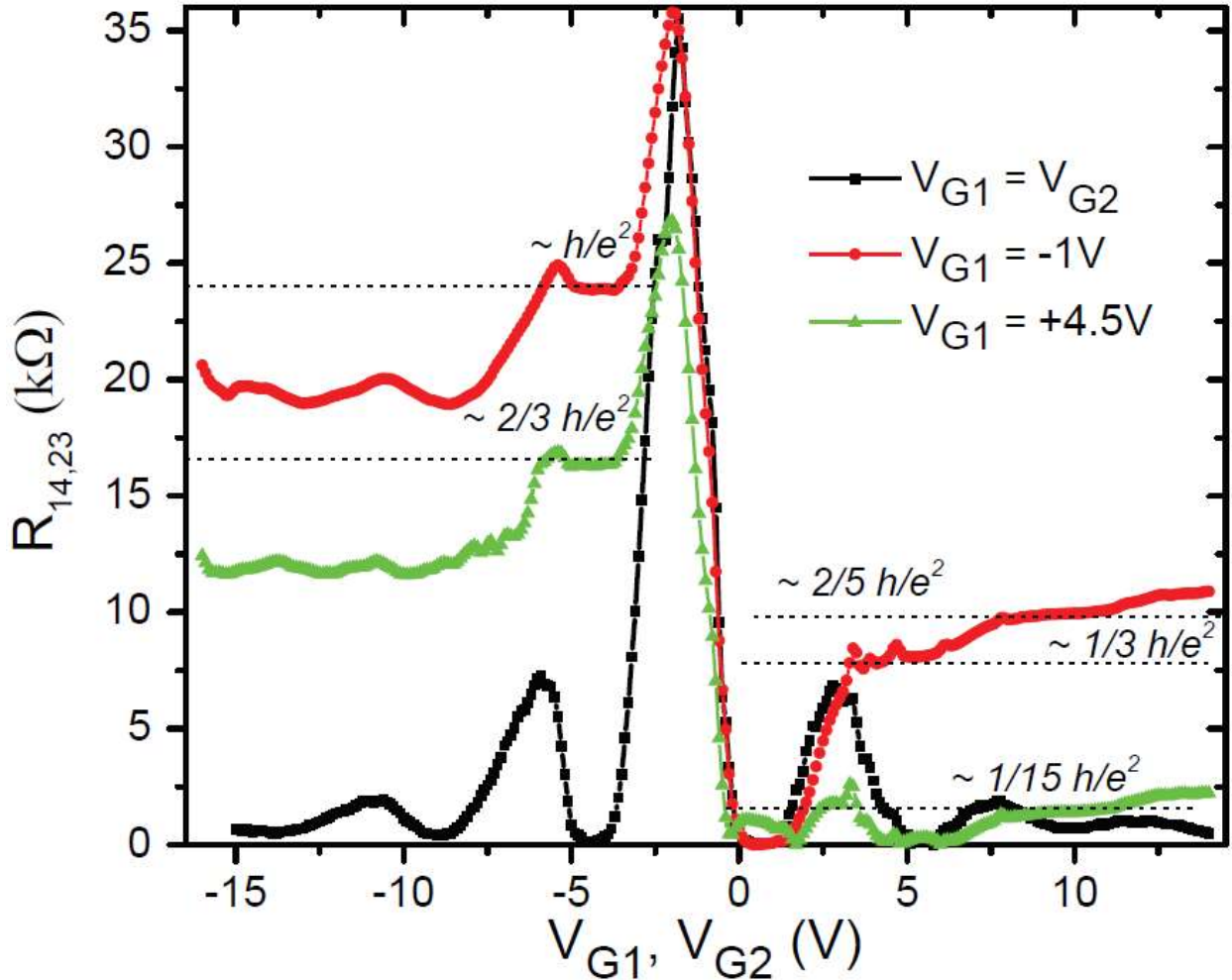

**Figure 1-SM.** A set of quantum Hall transport measurements were performed to collect data that would verify device functionality. The data are similar to those seen in similar devices [1]. In this example, a diagonal profile and two profiles with a fixed gate voltage in one half of the device are shown. The magnetic field is set to 12 T.

REFERENCES

[1] N. N. Klimov, S. T. Le, J. Yan, P. Agnihotri, E. Comfort, J. U. Lee, D. B. Newell, and C. A. Richter, Phys. Rev. B 92, 241301 (2015).